\documentclass[aps,pre,reprint]{revtex4-2}
\usepackage{amsmath,amssymb}
\usepackage{physics}
\usepackage{hyperref}
\usepackage{graphicx}

\begin{document}

\title{Electron Coherence and Dephasing by Quasi-Static Coulomb Fluctuations in Classical Ionic Media}

\author{Yury A. Budkov}
\email{ybudkov@hse.ru}
\affiliation{Laboratory of Computational Physics, HSE University, Tallinskaya st. 34, 123458 Moscow, Russia}
\affiliation{Laboratory of Multiscale Modeling of Molecular Systems, G.A. Krestov Institute of Solution Chemistry of the Russian Academy of Sciences, Ivanovo, Russia}

\begin{abstract}
We develop a microscopic theory of quantum-particle propagation through quasi-static electrostatic fluctuations of a classical one-component plasma. The RPA potential is Gaussian but has an unscreened $1/r$ correlation tail, which makes the disorder-averaged retarded Green function infrared sensitive. Its eikonal attenuation scale is $\ell_{\rm w}=k^2/G$ for $g=G/k^3\ll1$, with the formal strong-saddle asymptote $\ell_{\rm s}=(4\sqrt[3]{2}/3)G^{-1/3}$ for $g\gg1$. For a slab of thickness $d$, the retarded--advanced mutual-coherence kernel is infrared finite: the coherence decays algebraically for $\lambda_D\ll\rho\ll d$ and saturates at larger transverse separation. Thus equilibrium ionic-density fluctuations constitute a distinct quasi-static dephasing channel with a nonexponential coherence law. A finite coherence radius exists only when the saturation threshold is crossed. Relativistic paraxial reduction gives $a_{\rm rel}=1/(\hbar v)$ and the finite high-energy limit $1/(\hbar c)$. The ionic contribution is very small at conventional TEM energies for a $100\,\mathrm{nm}$ water-like layer. Within the quasi-static model it grows for thicker samples and lower-energy electrons, while sub-keV transmission through liquid water is strongly limited by inelastic scattering.
\end{abstract}

\maketitle

\section{Introduction}

Disorder-induced attenuation and Anderson localization are central themes in the theory of quantum transport~\cite{anderson1958absence,lifshits1988introduction,kramer1993localization,lee1985disordered}. Most microscopic treatments assume short-range impurities or a random potential with a finite correlation length. Coulomb systems are qualitatively different: charge-density fluctuations can generate a slowly decaying potential correlator, so attenuation of an averaged propagator, phase decorrelation, and genuine localization need not be governed by the same scale. We show below that equilibrium ionic-density fluctuations provide a distinct quasi-static dephasing channel, with long-range Coulomb correlations producing algebraic mutual-coherence decay over an intermediate range of transverse separations.

Kramer and MacKinnon~\cite{kramer1993localization} reviewed weak and strong localization and the scaling theory of the Anderson transition. Bergmann~\cite{bergman1984weak} discussed weak-localization corrections in thin metallic films, while Lee and Ramakrishnan~\cite{lee1985disordered} developed the connection between disorder, interference, and electron--electron interactions. Scaling approaches to electron transport in disordered conductors can relate the localization length directly to the correlation properties of the random potential~\cite{Satanin1991}, while correlated random potentials more generally can produce nonstandard propagation and localization behavior, as reviewed by Izrailev, Krokhin, and Makarov~\cite{izrailev2012anomalous}. Long-range Coulomb disorder also enters the Efros--Shklovskii description of disordered electronic systems~\cite{shklovskii2013electronic,shklovskii2024half}. The related effects of long-range interactions and disorder have been considered for ultracold plasmas and quantum Coulomb glasses~\cite{sous2019many,Vojta1998quantum}.

The present work addresses a different question from these approaches to disordered quantum systems and localization. We start from equilibrium charge fluctuations of a classical ionic medium and derive microscopically the Coulomb random-potential correlator seen by an external quantum particle. We then separate two observables that need not be governed by the same scale: attenuation of the disorder-averaged one-particle propagator and loss of mutual coherence between two transmitted amplitudes. The former is infrared sensitive because the potential correlator retains a $1/r$ tail, whereas in the latter the common long-wavelength phase cancels. This cancellation produces an infrared-finite finite-slab coherence law, including an algebraic intermediate regime and a saturation criterion for the existence of a finite coherence radius.

The ionic OCP used below is an idealized limit of systems in which one ionic component is mobile on the time scale of interest while the
compensating charge is effectively frozen. At a coarse-grained level, this separation is approached in superionic conductors, where a mobile
ionic sublattice coexists with a comparatively rigid charge-compensating framework~\cite{Rettie2021}, and in ion-exchange or polyelectrolyte membranes with fixed charged groups and mobile
counterions~\cite{Doyle2001,Spry2007}. The uniform OCP background represents only the long-wavelength charge compensation. Lattice periodicity, chemical binding, and structural disorder of the host are not included. The one-component restriction is not essential at Gaussian level: as shown explicitly below, a neutral multicomponent electrolyte gives the same long-wavelength random-potential correlator in the Debye--H\"uckel approximation~\cite{budkov2024statistical}, with $\kappa$ equal to the total screening wave number.

\begin{figure*}[t]
\centering
\includegraphics[width=0.98\textwidth]{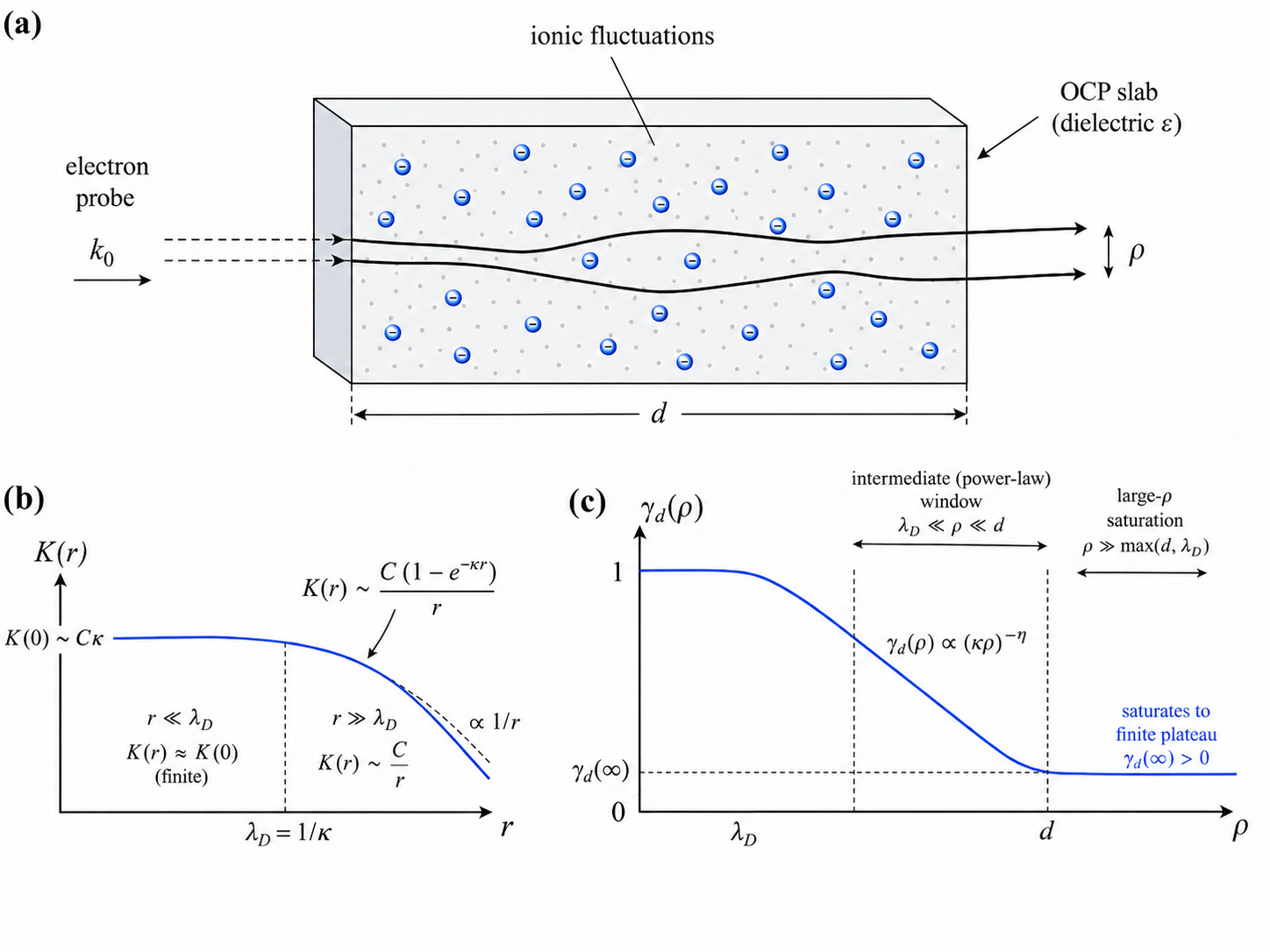}
\caption{Physical picture of fluctuation-induced attenuation and mutual-coherence degradation in an ionic one-component plasma. (a) An electron probe traverses a slab of thickness $d$ containing one mobile ionic species in a uniform neutralizing background. Two nearby paraxial trajectories separated by the transverse distance $\rho$ accumulate different random Coulomb phases. (b) The RPA potential correlator $K(r)=C(1-e^{-\kappa r})/r$ is finite at the origin and crosses over on the Debye scale $\lambda_D=\kappa^{-1}$ to the long-range Coulomb tail $K(r)\sim C/r$. (c) The finite-slab mutual coherence $\gamma_d(\rho)$ exhibits an algebraic intermediate regime, $\gamma_d(\rho)\propto(\kappa\rho)^{-\eta}$ for $\lambda_D\ll\rho\ll d$, and approaches a nonzero plateau at large $\rho$. The one-particle attenuation scale $\ell_{\rm w}$ is shown separately from the transverse coherence scales.}
\label{fig:physical_picture}
\end{figure*}

Figure~\ref{fig:physical_picture} summarizes the physical picture before the formal derivation. In panel~(a), an external electron crosses an ion-containing slab along an approximately straight paraxial path. Two nearby coherent amplitudes, separated by a transverse distance $\rho$, sample the same quasi-static ionic configuration but accumulate slightly different Coulomb phases. Panel~(b) shows the key statistical property of this environment: Debye screening makes the random-potential correlator finite at short distance, yet the correlator retains a long-range $1/r$ tail beyond $\lambda_D$. Consequently, a single disorder-averaged amplitude is sensitive to very long wavelength fluctuations. In the two-ray observable of panel~(c), however, the common part of the random phase cancels. The resulting mutual coherence is therefore infrared finite, decays algebraically over the intermediate range $\lambda_D\ll\rho\ll d$, and saturates when the transverse separation exceeds the finite-slab crossover scale. The one-particle attenuation length and the transverse coherence scales should thus be viewed as physically distinct quantities.

Efimov developed a nonperturbative representation of the disorder-averaged Green function~\cite{efimov2015quantum}. Here we combine that path-integral construction with the statistical mechanics
of a classical OCP\@. A quantum test particle interacts with the instantaneous random potential energy $W=q_0\phi$ generated by equilibrium ionic charge-density fluctuations. Within the random phase approximation (RPA), $W$ is a Gaussian field with the correlation
function
\begin{equation}\label{eq:K}
K(r)\equiv
\left\langle W(\mathbf r)W(\mathbf 0)\right\rangle
=
\frac{k_{\mathrm B}Tq_0^2}{4\pi\varepsilon}
\frac{1-e^{-\kappa r}}{r}
\equiv
C\frac{1-e^{-\kappa r}}{r},
\end{equation}
where $\kappa$ is the inverse Debye length. The correlator is finite at the origin but approaches $C/r$ at large distances. Its longitudinal line integral is therefore logarithmically infrared divergent. The disorder-averaged retarded Green function is treated in the eikonal approximation. Its exponential decay defines the \emph{eikonal attenuation length}. The saddle-point regimes are controlled by the dimensionless parameter $g=G/k^3$, where $\hbar k$ is the quantum-particle momentum, whereas $k/\kappa$ independently controls the validity of the straight-line approximation.

For a pair of paraxial rays crossing a slab of finite thickness, relevant to liquid-cell electron microscopy and diffraction~\cite{deJonge2011,Keskin2016,Cowley1995,Egerton2011,deJonge2019}, we use a normalized retarded--advanced kernel. The common long-wavelength phase cancels between the two paths, leaving an infrared-finite structure function. The coherence is algebraic for $\lambda_D\ll\rho\ll d$ and saturates at larger transverse separation. A relativistic paraxial reduction is used for the electron-beam estimates.

\section{Model of the plasma}
\label{sec:model}
We consider a one-component plasma (OCP) consisting of mobile ions of charge $q$ and a uniform neutralizing background that ensures overall electroneutrality. The ions are treated as classical point particles with average number density $n_0$ and local number density $n(\mathbf{r})$. The background is static and does not contribute to the fluctuating potential. Throughout the paper, $c$ denotes the speed of light. The medium is characterized by the absolute permittivity $\varepsilon=\varepsilon_0\varepsilon_r$, where $\varepsilon_r$ accounts for solvent screening or background polarizability. SI electromagnetic units are used throughout.

Here, “plasma” denotes a classical ionic Coulomb subsystem, not necessarily a gaseous plasma. The neutralizing background may represent either the conventional uniform electronic background or, at a coarse-grained level, an effectively immobile oppositely charged matrix. In the latter interpretation, the model applies on length scales large compared with the microscopic structure of the matrix. Accordingly, lattice periodicity, chemical binding, and quenched structural disorder are not resolved within the present simplified model.

A quantum test particle of charge $q_0$ interacts with the instantaneous electrostatic potential created by the fluctuating ion density:
\begin{equation}\label{eq:pot}
W(\mathbf{r}) = q_0 \int d^3r' \,
\frac{q\,\delta n(\mathbf{r}')}{4\pi\varepsilon|\mathbf{r} - \mathbf{r}'|},
\end{equation}
where $\delta n(\mathbf{r}) = n(\mathbf{r}) - n_0$ is the fluctuation of the ion concentration away from its mean value, and $n(\mathbf{r})$ is the local ion number density.
The charge density fluctuation is $\delta\rho_c(\mathbf{r}) = q\,\delta n(\mathbf{r})$ with $\langle \delta\rho_c(\mathbf{r})\rangle = 0$.

\section{Correlation function of the random potential}
\label{sec:correlation}
This section supplies the statistical input required for the propagation theory. Starting from Gaussian density fluctuations of the OCP, we derive the two-point correlator of the random potential experienced by the quantum particle. The central result is that Debye screening regularizes this correlator at short distances but does not remove its long-distance $1/r$ tail; this distinction is the origin of the infrared behavior encountered below.

To obtain the statistics of $W(\mathbf{r})$, we consider the effective fluctuation Hamiltonian of the classical one-component ion plasma~\cite{budkov2024statistical}.
Expanding the ideal-gas free energy to quadratic order in the density fluctuation $\delta n(\mathbf{r})=n(\mathbf{r})-n_0$ and measuring the free energy in units of $\beta^{-1}=k_{\mathrm B}T$ gives
\begin{equation*}
\beta\mathcal{F}_{\mathrm{id}} \simeq
\frac{1}{2}\int d^3r\,\frac{\bigl(\delta n(\mathbf{r})\bigr)^{2}}{n_0}.
\end{equation*}
The Coulomb energy in a dielectric medium is
\begin{equation*}
\beta U_{\mathrm{C}} =
\frac{\beta}{2}\int d^3r\,d^3r'\,
\frac{q^{2}\,\delta n(\mathbf{r})\delta n(\mathbf{r}')}
{4\pi\varepsilon|\mathbf{r} - \mathbf{r}'|}.
\end{equation*}
Introducing the Fourier representation, the total action decouples into independent modes~\cite{borue1988statistical},
\begin{equation*}
\begin{aligned}
S={}&\frac{1}{2}\int\frac{d^{3}Q}{(2\pi)^{3}}\,
\biggl(\frac{1}{n_0}+\frac{\beta q^{2}}{\varepsilon Q^{2}}\biggr)\bigl|\delta n(\mathbf{Q})\bigr|^{2}.
\end{aligned}
\end{equation*}
Hence $\delta n(\mathbf{Q})$ is a Gaussian field with variance
\begin{equation}\label{eq:var_OCP}
\begin{aligned}
\langle |\delta n(\mathbf{Q})|^{2}\rangle
&=\frac{1}{\frac{1}{n_0}+\frac{\beta q^{2}}{\varepsilon Q^{2}}}=\frac{\varepsilon k_{\mathrm B}T}{q^2}\,
\frac{\kappa^{2}Q^{2}}{Q^{2}+\kappa^{2}},
\end{aligned}
\end{equation}
where the inverse Debye screening length is $\kappa^{2}=n_0 q^{2}/(\varepsilon k_{\mathrm{B}}T)$.
Equation~(\ref{eq:var_OCP}) is the standard RPA result for the charge structure factor of a one-component plasma~\cite{barrat2003basic}.

From Eq.~\eqref{eq:pot}, the Fourier transform of the potential is $W(\mathbf{Q}) = q_0q\,\delta n(\mathbf{Q})/(\varepsilon Q^{2})$.
The resulting field $W$ is Gaussian, has zero mean, and has pair correlator
\begin{equation}\label{eq:corrk_OCP}
\begin{aligned}
\langle W(\mathbf{Q})W(-\mathbf{Q})\rangle
&=\Bigl(\frac{q_0 q}{\varepsilon Q^{2}}\Bigr)^{\!2}
\langle |\delta n(\mathbf{Q})|^{2}\rangle\\
&=\frac{q_0^{2} k_{\mathrm B}T}{\varepsilon}\,
\frac{\kappa^{2}}{Q^{2}(Q^{2}+\kappa^{2})}.
\end{aligned}
\end{equation}
Fourier inversion gives the coordinate-space correlation function
\begin{equation}\label{eq:K_OCP}
\begin{aligned}
K(r)&\equiv\langle W(\mathbf r)W(0)\rangle\\
&=\frac{q_0^{2}k_{\mathrm B}T}{\varepsilon}
\int\frac{d^{3}Q}{(2\pi)^{3}}\,
e^{i\mathbf Q\cdot\mathbf r}
\Bigl(\frac{1}{Q^{2}}-\frac{1}{Q^{2}+\kappa^{2}}\Bigr)\\
&=\frac{k_{\mathrm B}Tq_0^{2}}{4\pi\varepsilon}\,
\frac{1-e^{-\kappa r}}{r}.
\end{aligned}
\end{equation}
The two limiting forms of Eq.~\eqref{eq:K_OCP} are $K(r)\simeq k_{\mathrm B}Tq_0^2\kappa/(4\pi\varepsilon)$ for $r\ll\kappa^{-1}$ and $K(r)\simeq k_{\mathrm B}Tq_0^2/(4\pi\varepsilon r)$ for $r\gg\kappa^{-1}$. Thus, the potential fluctuations retain the long-range $1/r$ tail characteristic of three-dimensional Coulomb systems.

The Gaussian RPA is controlled for a weakly coupled OCP\@. Introducing the Bjerrum length $\ell_B=q^2/(4\pi\varepsilon k_{\mathrm B}T)$ and the Wigner--Seitz radius $a_{\rm WS}=(3/4\pi n_0)^{1/3}$, the conventional plasma parameter is $\Gamma_{\rm OCP}={\ell_B}/{a_{\rm WS}}$. The quadratic theory is quantitatively reliable for $\Gamma_{\rm OCP}\ll1$ and should be regarded as an RPA-level estimate when the coupling approaches unity. At the same Gaussian level, the potential correlator of a neutral multicomponent electrolyte has the same form as Eq.~\eqref{eq:corrk_OCP}, with $\kappa^2=\sum_{\alpha}n_{\alpha}q_{\alpha}^{2}/(\varepsilon k_{\mathrm B}T)$. Thus the finite-thickness formulas below carry over to the Debye--H\"uckel limit of a multicomponent electrolyte after replacing $\kappa$ by the total screening wave number. In particular, a symmetric $1{:}1$ electrolyte is covered at this Gaussian level even though both ionic species fluctuate. This equivalence concerns the static two-point potential correlator. Non-Gaussian correlations, finite ion size, and species-dependent dynamics are outside the present approximation.

\section{Efimov's path-integral formalism and the eikonal attenuation scale}
\label{sec:localization}

This section asks how the quasi-static ionic potential modifies a single coherent propagation amplitude. We first average Efimov's path-integral representation over the Gaussian ionic fluctuations and then specialize to a nearly rectilinear eikonal trajectory. The long-range tail of the potential correlator makes the resulting one-particle attenuation scale infrared sensitive; after introducing a physical longitudinal cutoff, the weak and formal strong saddle-point regimes follow from the dimensionless parameter $g$. We finally check when transverse fluctuations about the straight path remain parametrically small.

For a nonrelativistic quantum particle of mass $m$ in a static random potential $W(\mathbf{x})$, the retarded Green function at energy $E=\hbar^{2}k^{2}/(2m)$ may be written in Efimov's path-integral representation. Starting from the time representation of the resolvent and introducing the dimensionless variable $s=t\hbar k/(mr)$, with $r=|\mathbf{x}-\mathbf{x}'|$, one obtains~\cite{efimov2015quantum}
\begin{equation}\label{eq:G0_general}
\begin{aligned}
G_k(\mathbf{x},\mathbf{x}'|W)
={}&B\int_0^\infty\frac{ds}{s^{3/2}}
\int\frac{\mathcal D\boldsymbol\eta}{\mathcal C}\exp\left[\frac{i}{2}kr\left(s+\frac{1}{s}\right)\right]\\
&\times\exp\left[\frac{i}{2}\int_0^rdu\,
\dot{\boldsymbol\eta}^{2}(u)\right]\\
&\times\exp\left[-i\frac{ms}{k\hbar^2}\int_0^rdu\,
W\!\left(\mathbf{x}(u)\right)\right].
\end{aligned}
\end{equation}
where $\mathbf n=(\mathbf{x}-\mathbf{x}')/r$, $\mathbf{x}(u)=\mathbf{x}'+\mathbf n u-\sqrt{s/k}\,\boldsymbol\eta(u)$, and $\boldsymbol\eta(0)=\boldsymbol\eta(r)=0$. The constants $B$ and $\mathcal C$ normalize the representation and do not enter the decay exponent.

Averaging over the Gaussian disorder gives
\begin{equation}\label{eq:G_avg_general}
\begin{aligned}
\left\langle G_k(r)\right\rangle_W
={}&B\int_0^\infty\frac{ds}{s^{3/2}}
\int\frac{\mathcal D\boldsymbol\eta}{\mathcal C}\exp\left[\frac{i}{2}kr\left(s+\frac{1}{s}\right)\right]\\
&\times\exp\left[\frac{i}{2}\int_0^rdu\,
\dot{\boldsymbol\eta}^{2}(u)\right]\\
&\times\exp\left[-\frac{m^2s^2}{2\hbar^4k^2}
\int_0^rdu\int_0^rdu'\,
K\!\left(\mathbf n(u-u')\right.\right.\\
&\hspace{4.2em}\left.\left.{}-\sqrt{\frac{s}{k}}
[\boldsymbol\eta(u)-\boldsymbol\eta(u')]\right)\right].
\end{aligned}
\end{equation}
In the straight-line approximation, $\boldsymbol\eta=0$, the longitudinal double integral has the exact finite-distance form
\begin{equation}\label{eq:double_integral_exact}
\begin{aligned}
\int_0^rdu\int_0^rdu'\,K\!\left(|u-u'|\right)
&=2\int_0^rdu\,(r-u)K(u)\\
&=2Cr\,\mathcal H(\kappa r),
\end{aligned}
\end{equation}
where
\begin{equation}\label{eq:H_exact}
\mathcal H(x)=\ln x+\gamma_{\rm E}+E_1(x)-1+\frac{1-e^{-x}}{x},
\end{equation}
and $E_1(x)=\int_x^\infty dt\,e^{-t}/t$. Since $\mathcal H(x)\simeq\ln x+\gamma_{\rm E}-1$ for $x\gg1$, the ideal infinite static OCP does not possess an $r$-independent attenuation rate: the effective rate retains a logarithmic dependence on the propagation distance.

A conventional exponential attenuation scale can nevertheless be defined when a physical mechanism cuts off longitudinal correlations beyond a scale $L_{\rm IR}$ and the propagation distance satisfies $r\gg L_{\rm IR}$. Then the longitudinal double integral is $2r\int_0^{L_{\rm IR}}du\,K(u)$ to leading order, and the exponent per unit propagation length becomes
\begin{equation}\label{eq:Phi_G}
\begin{aligned}
\Phi(s)&=\frac{ik}{2}\left(s+\frac{1}{s}\right)-\frac{G}{k^2}s^2,\\
G&=\frac{m^2}{\hbar^4}\int_0^{L_{\rm IR}}du\,K(u).
\end{aligned}
\end{equation}
The cutoff may be supplied by dynamical decorrelation, a mean free path, a phase-coherence length, or an independent finite transverse size. The static bulk OCP model itself does not determine it. For the RPA correlator,
\begin{equation}\label{eq:G_exact}
G=\frac{m^2C}{\hbar^4}
\left(\ln(\kappa L_{\rm IR})+\gamma_{\rm E}+E_1(\kappa L_{\rm IR})\right),
\end{equation}
which reduces for $\kappa L_{\rm IR}\gg1$ to $G\simeq{m^2C}\ln(\kappa L_{\rm IR})/{\hbar^4}$, up to an additive constant under the logarithm. If no independent cutoff is present, the same saddle formulas may be used only as finite-distance estimates after replacing $G$ by $G_r=(m^2C/\hbar^4)\mathcal H(\kappa r)$.

The attenuation rate of the averaged propagator is $\Gamma(k)=\ell^{-1}(k)=-\operatorname{Re}\Phi(s_c)$, where the saddle point satisfies
\begin{equation}\label{eq:saddle_dimensionless}
\frac{i}{2}\left(1-\frac{1}{s_c^2}\right)-2g s_c=0,
\qquad
g=\frac{G}{k^3}.
\end{equation}
Thus the distinction between the weak and strong saddles is controlled by $g$, whereas $k/\kappa$ characterizes the separation between the particle wavelength and the Debye crossover length.

For $g\ll1$, the saddle continuously connected to the free-particle saddle is $s_c=1-2ig+O(g^2)$, and
\begin{equation}\label{eq:ell_weak}
\ell_{\rm w}(k)=\frac{k^2}{G}
\simeq
\frac{4\pi\varepsilon\hbar^4k^2}
{m^2k_{\mathrm B}Tq_0^2\ln(\kappa L_{\rm IR})}.
\end{equation}
For $g\gg1$, the same saddle approaches $s_c=(4g)^{-1/3}e^{-i\pi/6}+O(g^{-1})$ and gives the formal asymptotic estimate
\begin{equation}\label{eq:ell_strong}
\begin{aligned}
\ell_{\rm s}(k)
&=\frac{4\sqrt[3]{2}}{3}\,G^{-1/3}\simeq\frac{4\sqrt[3]{2}}{3}
\left(
\frac{4\pi\varepsilon\hbar^4}
{m^2k_{\mathrm B}Tq_0^2\ln(\kappa L_{\rm IR})}
\right)^{1/3}.
\end{aligned}
\end{equation}
The momentum independence of Eq.~\eqref{eq:ell_strong} is specific to the formal strong saddle. Its quantitative use still requires the straight-line approximation: if $g\gg1$ is reached by lowering $k$ to $k\lesssim\kappa$, transverse path fluctuations are no longer parametrically suppressed and Eq.~\eqref{eq:ell_strong} remains a scaling estimate.

The transverse stability of the rectilinear-path approximation can be checked explicitly. Appendix~\ref{app:transverse_fluctuations} evaluates the Gaussian determinant of the two fluctuations orthogonal to $\mathbf n$. In the weak saddle it gives
$\delta\Gamma_\perp/\Gamma_{\rm w}\simeq
0.59086\,(\kappa/k)g/\mathcal L^2$, where
$\mathcal L=\ln(\kappa L_{\rm IR})+\gamma_{\rm E}+E_1(\kappa L_{\rm IR})$.
Thus the leading transverse correction is parametrically small when both
$g\ll1$ and $\kappa/k\ll1$. The same calculation does not extend the
formal strong saddle into the region $k\lesssim\kappa$; rather, it makes
explicit why the straight-line expansion loses control there.

The infrared sensitivity of $\ell$ follows from the phase accumulated from long-wavelength potential fluctuations, including modes that are nearly uniform over the propagation region. A mutual-coherence observable depends on a phase difference and cancels this common contribution. Accordingly, $\ell$ characterizes attenuation of the ensemble-averaged propagator. It is distinct from a typical Anderson localization length or a transport mean free path~\cite{abrikosov1988fundamentals}.

It is worth noting that the logarithm in Eq.~\eqref{eq:G_exact} resembles the Coulomb logarithm of plasma kinetic theory~\cite{pitaevskii2012physical}, but its origin is different. In kinetic theory the logarithm follows from an impact-parameter integral over scattering events. Here it is generated directly by the $1/r$ tail of the equilibrium potential correlator and is regularized only by the large-distance scale $L_{\rm IR}$.

\section{Mutual-coherence propagator}
\label{sec:coherence}

This section turns from attenuation of a single averaged amplitude to the experimentally more direct question of mutual coherence between two nearby transmitted amplitudes. The relevant quantity is a phase difference, so sufficiently long-wavelength fluctuations that are common to both paths cancel. We derive the corresponding retarded--advanced representation, obtain the finite-slab phase structure function for two parallel rays, and then identify its algebraic intermediate regime, its large-separation saturation, and the criterion for the existence of a finite coherence radius.

\subsection{Retarded--advanced representation}

In a coherent electron-optical experiment the contrast is controlled by correlations between complex amplitudes propagated along two nearby paraxial trajectories. For general initial and final points, we define the retarded--advanced kernel
\begin{equation*}
\Gamma_{12}=\left\langle G^R(\mathbf R_1,\mathbf r_1)G^A(\mathbf R_2,\mathbf r_2)\right\rangle.
\end{equation*}
This retarded--advanced object is the standard type of two-point kernel used in mesoscopic propagation theory~\cite{akkermans2007mesoscopic,abrikosov1988fundamentals}. We use it only as a mutual-coherence kernel. For two statistically equivalent trajectories, the normalized degree of coherence is $\gamma_{12}=\Gamma_{12}/\sqrt{\Gamma_{11}\Gamma_{22}}$. In the parallel-ray geometry considered below, $\gamma_{12}=\gamma_d(\rho)$ depends only on the fixed transverse separation $\rho$. Because the ionic configuration is treated as quasi-static, this quantity describes ensemble-averaged phase decorrelation. Dynamical inelastic decoherence is not included.

Using Eq.~\eqref{eq:G0_general} for the retarded and advanced propagators gives
\begin{equation*}
\begin{aligned}
\Gamma_{12}={}&B^2\int\frac{ds_1ds_2}{(s_1s_2)^{3/2}}e^{i\Phi_0}
\int\mathcal D\boldsymbol\eta_1\mathcal D\boldsymbol\eta_2\\
&\times\exp\left[iS_0[\boldsymbol\eta_1]-iS_0[\boldsymbol\eta_2]\right]\\
&\times\left\langle e^{-iS_{W,1}+iS_{W,2}}\right\rangle.
\end{aligned}
\end{equation*}
Here $\Phi_0=\frac{k}{2}\{d_1(s_1+s_1^{-1})-d_2(s_2+s_2^{-1})\}$. The path actions are $S_0[\boldsymbol\eta_i]=\frac12\int_0^{d_i}du\,\dot{\boldsymbol\eta}_i^2(u)$ and $S_{W,i}=(ms_i/k\hbar^2)\int_0^{d_i}du\,W(\mathbf r_i(u))$. For Gaussian disorder,
\begin{equation*}
\left\langle e^{-iS_{W,1}+iS_{W,2}}\right\rangle
=\exp\left[-\frac12\left\langle(S_{W,1}-S_{W,2})^2\right\rangle\right].
\end{equation*}
In the weak-attenuation eikonal regime the longitudinal saddles remain close to $s_1=s_2=1$, and the nonrelativistic phase coupling is $a=m/(\hbar^2k)=1/(\hbar v)$.

\subsection{Finite-thickness structure function in the bulk-correlator approximation}

We consider two parallel straight rays traversing a slab of thickness $d$ at fixed transverse separation $\boldsymbol\rho$. The correlator inside the slab is taken to be translationally invariant; dielectric-boundary and image-charge corrections are neglected. Within this straight-ray bulk-correlator approximation, the normalized mutual coherence has the Gaussian form
\begin{equation}\label{eq:gamma_finite}
\gamma_d(\rho)=\exp\left(-\frac{1}{2}\mathcal D_d(\rho)\right),
\end{equation}
where the phase structure function is
\begin{equation*}
\begin{aligned}
\mathcal D_d(\rho)
={}&2a^2\int_0^d dz\int_0^d dz'\\
&\times\left[
K(|z-z'|)-K\!\left(\sqrt{(z-z')^2+\rho^2}\right)
\right].
\end{aligned}
\end{equation*}
Equivalently,
\begin{equation}\label{eq:Dphi_finite}
\mathcal D_d(\rho)
=4a^2\int_0^d du\,(d-u)
\left(
K(u)-K\!\left(\sqrt{u^2+\rho^2}\right)
\right).
\end{equation}
Equation~\eqref{eq:Dphi_finite}, rather than the infinite-path approximation, is the appropriate starting point for a finite liquid cell.

For $d\gg\max(\kappa^{-1},\rho)$, one may replace $d-u$ by $d$ over the effective range of the difference and extend the upper limit to infinity. Substitution of Eq.~\eqref{eq:K} then gives
\begin{equation}\label{eq:Dphi_thick}
\mathcal D_d(\rho)
\simeq4a^2Cd\,\mathcal F(\kappa\rho),
\end{equation}
where
\begin{equation}\label{eq:F_exact}
\mathcal F(x)
=\int_0^xdt\left(\frac{1}{t}-K_1(t)\right)
=\ln\frac{x}{2}+\gamma_{\rm E}+K_0(x).
\end{equation}
Here $K_0$ and $K_1$ are modified Bessel functions.

At small transverse separation, $\mathcal F(x)=\frac{x^2}{4}[\ln(2/x)+1-\gamma_{\rm E}]+O(x^4\ln x)$, so $\mathcal D_d(\rho)\simeq a^2Cd\,\kappa^2\rho^2[\ln(2/\kappa\rho)+1-\gamma_{\rm E}]$. If an $e^{-1}$ coherence radius $\rho_c$ exists and obeys both $\kappa\rho_c\ll1$ and $\rho_c\ll d$, then in the weak-attenuation regime
\begin{equation}\label{eq:rhoc_conditional}
\begin{aligned}
\rho_c^2
\left(\ln\frac{2}{\kappa\rho_c}+1-\gamma_{\rm E}\right)
&=\frac{2}{a^2Cd\,\kappa^2},\\
\rho_c&\sim\lambda_D\sqrt{\frac{\ell_{\rm w}}{d}}.
\end{aligned}
\end{equation}
where the last relation suppresses the logarithmic and order-unity factors shown explicitly in the preceding relation. Equation~\eqref{eq:rhoc_conditional} is therefore only a conditional scaling estimate: it applies to the weak saddle with $\kappa\rho_c\ll1$ and only when the finite-slab structure function reaches the value $2$.

For $\lambda_D\ll\rho\ll d$, $\mathcal D_d(\rho)\simeq4a^2Cd[\ln(\kappa\rho/2)+\gamma_{\rm E}]$, and therefore
\begin{equation}\label{eq:power_law}
\gamma_d(\rho)\propto(\kappa\rho)^{-\eta},
\qquad
\eta=2a^2Cd.
\end{equation}
The power law is an intermediate asymptotic. It does not persist to arbitrary transverse separation in a finite slab, in direct analogy with structure-function crossovers in long-range-correlated random media~\cite{tatarskii1971effects,ishimaru1997wave}.

\subsection{Saturation and the existence criterion for a coherence radius}

The finite slab changes the large-separation behavior qualitatively: once the two rays are sufficiently far apart, increasing $\rho$ no longer produces an unbounded growth of the phase-difference variance. We now evaluate this saturation plateau and use it to determine whether the mutual coherence ever reaches the conventional $e^{-1}$ threshold at a finite transverse separation.

Taking $\rho\to\infty$ directly in Eq.~\eqref{eq:Dphi_finite} gives $\mathcal D_d(\infty)=4a^2\int_0^d du\,(d-u)K(u)$. For the OCP correlator,
\begin{equation}\label{eq:Dinf_OCP}
\begin{aligned}
\mathcal D_d(\infty)&=4a^2Cd\,\mathcal H(\kappa d),\\
\gamma_d(\infty)&=\exp\left[-2a^2Cd\,\mathcal H(\kappa d)\right].
\end{aligned}
\end{equation}
where $\mathcal H$ is defined in Eq.~\eqref{eq:H_exact}. The one-ray coherent factor is
\begin{equation*}
\left\langle e^{-ia\int_0^d dz\,W(z)}\right\rangle
=\exp\left[-a^2Cd\,\mathcal H(\kappa d)\right],
\end{equation*}
so the large-separation mutual coherence is its square, as expected for two statistically independent distant rays. For the OCP correlator, $K(r)$ is positive and strictly decreasing, so $\mathcal D_d(\rho)$ is continuous and monotonically increasing. A finite coherence radius therefore exists if and only if
\begin{equation}\label{eq:existence_condition}
\mathcal D_d(\infty)>2.
\end{equation}
The equality $\mathcal D_d(\infty)=2$ is marginal and corresponds to $\rho_c\to\infty$. When the inequality is not fulfilled, the ionic disorder causes a finite reduction of mutual coherence but does not define a finite transverse $e^{-1}$ decay length. This criterion relies on the monotonic OCP correlator and need not apply unchanged to an oscillatory nonlocal dielectric model~\cite{KornyshevSutmann1996}. In the weak-attenuation regime, Eqs.~\eqref{eq:ell_weak} and \eqref{eq:Dinf_OCP} provide the finite-thickness link
\begin{equation}\label{eq:Dinf_ell_relation}
\mathcal D_d(\infty)
=\frac{4d\,\mathcal H(\kappa d)}
{\ell_{\rm w}\ln(\kappa L_{\rm IR})},
\end{equation}
within logarithmic accuracy for the attenuation scale. Equation~\eqref{eq:Dinf_ell_relation} connects one-particle attenuation to mutual-coherence loss. A coherence radius follows from it only when the existence and small-separation conditions stated above are satisfied.

\section{Examples of applications and numerical estimates}
\label{sec:estimates}

\subsection{Physical realizations and observables}

In a superionic conductor or a fixed-charge ionic medium, the OCP can represent the mobile ionic subsystem after coarse-graining over the oppositely charged host. A liquid binary electrolyte contains fluctuations of both charge signs, but at Debye--H\"uckel level gives the same Gaussian random-potential correlator, so the finite-slab coherence formulas carry over with the total screening wave number.

For the electron-optical application, the quantum particle is an external probe with $q_0=-e$, where $e$ is the elementary charge. The function $\gamma_d(\rho)$ is the normalized mutual coherence of amplitudes propagated along two transversely separated paths. Its reduction contributes directly to the loss of electron-beam interference or diffraction contrast. The symbol $\ell$ retains the meaning defined in Sec.~\ref{sec:localization}: the attenuation scale of the disorder-averaged one-particle propagator.

\subsection{Regimes and physical cutoffs}

The saddle parameter $g=G/k^3$ and the straight-line condition constrain different parts of the approximation. A sufficient high-energy condition is $k\gg\kappa$, together with $kd\gg1$ for a finite slab. Thus $g$ can approach unity while $k/\kappa$ remains large, whereas lowering $k$ may spoil the eikonal expansion before the formal strong saddle becomes quantitative. The infrared cutoff $L_{\rm IR}$ need not coincide with the slab thickness $d$. In a finite static sample it is set by the longest longitudinal scale over which the potential remains correlated along a coherent trajectory. In a dynamical medium it may be bounded by scales such as $v\tau_{\rm ion}$, a dynamical dephasing or inelastic-scattering length, and, when angular deflection becomes important, the relevant transport mean free path; determining it self-consistently requires a dynamical transport theory. In practice, $L_{\rm IR}$ may also be regarded as an effective phenomenological parameter and, provided the eikonal attenuation is experimentally accessible and the other medium parameters are independently known, can in principle be extracted by fitting the measured attenuation scale. The finite-slab structure function in Eq.~\eqref{eq:Dphi_finite}, on the other hand, is infrared finite and contains $d$ explicitly. Equation~\eqref{eq:ell_strong} should therefore be read as the asymptote of the strong saddle within the straight-line construction. Transverse wandering, pinning, and an Anderson transition require information beyond that approximation, for example fluctuations about the straight path or a typical two-particle quantity.

For comparison with a liquid-cell electron experiment, the straight-ray criterion should be distinguished from survival in the zero-loss channel. The inelastic mean free path $\lambda_{\rm in}$ is the mean distance between inelastic events and therefore does not by itself test rectilinearity; zero-loss electrons may still undergo elastic angular scattering~\cite{Egerton2011}. Relativistic full-Penn calculations give $\lambda_{\rm in}\simeq159\,\mathrm{nm}$ for liquid water at $98.7\,\mathrm{keV}$~\cite{FloresMancera2020}, while direct EELS measurements at $120$ and $300\,\mathrm{keV}$ show that the experimentally extracted water IMFP depends on the electron-optical collection geometry~\cite{Yesibolati2020}. Thus, for the $100\,\mathrm{keV}$, $d=100\,\mathrm{nm}$ reference point used below, $\lambda_{\rm in}/d\simeq1.6$ and the simple Poisson no-inelastic-event fraction is $\exp(-d/\lambda_{\rm in})\simeq0.53$. For the same reference point, $\kappa/k\simeq4.3\times10^{-4}$ and $kd\simeq1.7\times10^{5}$, so the geometric high-energy conditions of the straight-ray expansion are well satisfied within the static model. The static coherence result should therefore be interpreted as applying to the near-forward paraxial component conditioned on the absence of inelastic events, rather than to a strictly ballistic beam. A quantitative treatment of inelastic decoherence would require the dynamical response of the medium and lies outside the present quasi-static theory.

\subsection{Electron-beam coherence: OCP reference and Debye--H\"uckel electrolyte limit}

For the numerical reference calculation we take monovalent mobile ions, $|q|=e$, and an electron probe, $q_0=-e$. Since the correlator depends on $q_0^2$, only $|q_0|=e$ enters the results. The OCP number density is chosen to correspond to $0.1\,\mathrm{mol\,L^{-1}}$ at $T=300\,\mathrm K$ and $\varepsilon_r=80$. Equation~\eqref{eq:var_OCP} gives $\lambda_D=\kappa^{-1}\simeq1.38\,\mathrm{nm}$. The corresponding coupling parameter is $\Gamma_{\rm OCP}\simeq0.44$, so the calculation is an RPA reference for a weak-to-moderately coupled ionic medium, not a precision model of an aqueous salt. A symmetric $1{:}1$ electrolyte at the same salt concentration has inverse screening length larger by $\sqrt{2}$ in the Debye--H\"uckel limit.

For a slab of thickness $d=100\,\mathrm{nm}$, the saturation formula~\eqref{eq:Dinf_OCP}, together with the relativistic coupling derived below, gives the values listed in Table~\ref{tab:finite_coherence}.
\begin{table}[t]
\caption{Finite-thickness mutual-coherence estimates for a
$0.1\,\mathrm{mol\,L^{-1}}$ OCP with $|q|=|q_0|=e$,
$T=300\,\mathrm K$, $\varepsilon_r=80$, and $d=100\,\mathrm{nm}$.}
\label{tab:finite_coherence}
\begin{ruledtabular}
\begin{tabular}{cccc}
kinetic energy & $v/c$ & $\mathcal D_d(\infty)$ & $\gamma_d(\infty)$ \\
$100\,\mathrm{keV}$ & $0.548$ & $6.2\times10^{-5}$ & $0.999969$ \\
$1\,\mathrm{keV}$ & $0.0625$ & $4.75\times10^{-3}$ & $0.99763$ \\
$100\,\mathrm{eV}$ & $0.0198$ & $4.74\times10^{-2}$ & $0.9766$ \\
$5\,\mathrm{eV}$ & $4.42\times10^{-3}$ & $0.948$ & $0.623$
\end{tabular}
\end{ruledtabular}
\end{table}
For all four energies, $\mathcal D_d(\infty)<2$, so the coherence stays above $e^{-1}$ at every transverse separation. The ionic contribution is negligible at $100\,\mathrm{keV}$ for a $100\,\mathrm{nm}$ layer and remains only a few percent at $100\,\mathrm{eV}$. At $5\,\mathrm{eV}$ the plateau falls to $\gamma_d(\infty)\simeq0.623$. The $1\,\mathrm{keV}$ and lower-energy entries should be read as formal trends of the quasi-static model rather than realistic zero-loss transmission conditions through $100\,\mathrm{nm}$ of liquid water: the same IMFP calculation gives $\lambda_{\rm in}\simeq3.94\,\mathrm{nm}$ at $0.99\,\mathrm{keV}$ and $\simeq1.01\,\mathrm{nm}$ at $99.5\,\mathrm{eV}$~\cite{FloresMancera2020}. Multiple scattering, inelastic losses, boundary effects, and ionic dynamics therefore become essential in this range.

For comparison, the symmetric $1{:}1$ Debye--H\"uckel electrolyte at the same salt concentration has $\kappa_{1:1}=\sqrt{2}\,\kappa_{\rm OCP}$ and $\lambda_D\simeq0.98\,\mathrm{nm}$. Substitution into Eq.~\eqref{eq:Dinf_OCP} gives $\gamma_d(\infty)\simeq0.999966$, $0.99742$, and $0.9745$ at $100\,\mathrm{keV}$, $1\,\mathrm{keV}$, and $100\,\mathrm{eV}$, respectively. Thus the conclusion drawn from the OCP reference calculation is unchanged at the Debye--H\"uckel level for a dilute symmetric electrolyte.

Increasing the OCP density from $0.1$ to $1\,\mathrm{mol\,L^{-1}}$ reduces the Debye length to approximately $0.436\,\mathrm{nm}$. For $d\gg\lambda_D$, the density dependence of the saturated structure function is only logarithmic through $\mathcal H(\kappa d)$. At $1\,\mathrm{mol\,L^{-1}}$ the formal RPA plateau values are approximately $0.999960$, $0.99693$, and $0.9698$ at $100\,\mathrm{keV}$, $1\,\mathrm{keV}$, and $100\,\mathrm{eV}$, respectively. At this density $\Gamma_{\rm OCP}\simeq0.95$, so point-ion RPA is no longer quantitatively controlled and these numbers illustrate only the model trend.

The thickness dependence is more pronounced because Eq.~\eqref{eq:Dinf_OCP} grows approximately as $d\ln(\kappa d)$. For the same $0.1\,\mathrm{mol\,L^{-1}}$ OCP, increasing the thickness to $1\,\mu\mathrm m$ gives $\gamma_d(\infty)\simeq0.99951$, $0.9629$, and $0.686$ at $100\,\mathrm{keV}$, $1\,\mathrm{keV}$, and $100\,\mathrm{eV}$, respectively. The static model thus predicts appreciably stronger phase decorrelation in thicker layers and for sub-keV electrons. In that range, elastic multiple scattering, energy loss, dielectric-boundary effects, and ionic dynamics must be included for a quantitative description.

\begin{figure*}[t]
\centering
\includegraphics[width=0.98\textwidth]{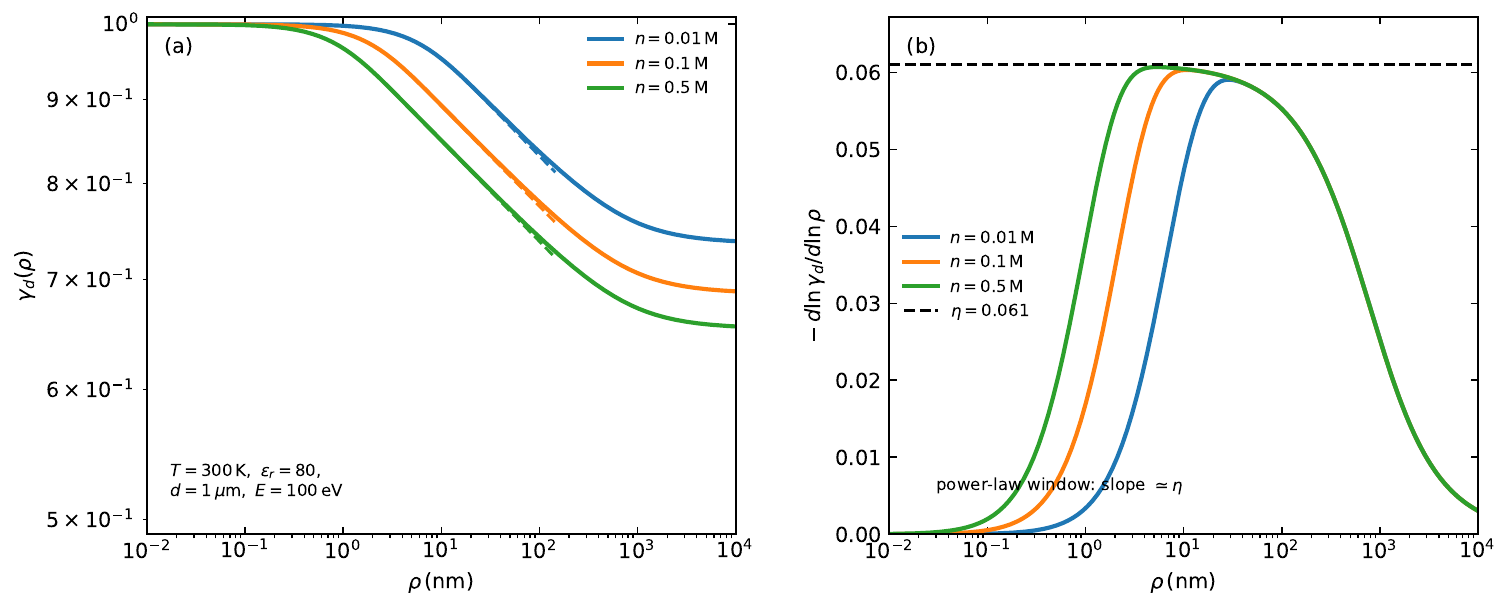}
\caption{Finite-thickness mutual coherence for a water-like OCP, computed from the exact slab formula~\eqref{eq:Dphi_finite}. The parameters are $T=300\,\mathrm K$, $\varepsilon_r=80$, $d=1\,\mu\mathrm m$, and $E=100\,\mathrm{eV}$, with concentrations $n=0.01$, $0.1$, and $0.5\,\mathrm{mol\,L^{-1}}$. (a) Exact coherence envelopes $\gamma_d(\rho)$ (solid) together with the intermediate-asymptotic power laws from Eq.~\eqref{eq:power_law} (dashed). (b) Local exponent $n_{\mathrm{eff}}(\rho)=-d\ln \gamma_d/d\ln \rho$. The plateau at $n_{\mathrm{eff}}\simeq \eta$ demonstrates the crossover to the algebraic regime $\lambda_D\ll \rho \ll d$, while the final decay of $n_{\mathrm{eff}}$ to zero shows the finite-thickness saturation of the coherence function.}
\label{fig:coherence_profiles}
\end{figure*}

Figure~\ref{fig:coherence_profiles} illustrates the full finite-thickness coherence envelope obtained from Eq.~\eqref{eq:Dphi_finite}. We choose a comparatively thick slab, $d=1\,\mu\mathrm m$, and a sub-keV electron energy, $E=100\,\mathrm{eV}$, in order to make the crossover structure visible within the present quasi-static model. Panel~(a) shows $\gamma_d(\rho)$ for three OCP concentrations at fixed $T=300\,\mathrm K$ and $\varepsilon_r=80$; the dashed segments represent the intermediate-asymptotic power law from Eq.~\eqref{eq:power_law}. Increasing the concentration decreases $\lambda_D$ and therefore shifts the onset of the algebraic window to smaller separations. Panel~(b) displays the local exponent $n_{\mathrm{eff}}(\rho)=-d\ln\gamma_d(\rho)/d\ln\rho$. The approach of $n_{\mathrm{eff}}$ to a plateau at $\eta$ provides a direct visualization of the power-law regime $\lambda_D\ll \rho \ll d$. When $\rho$ becomes comparable to the slab thickness, the exact finite-thickness result crosses over away from the power law and the local exponent returns toward zero, reflecting saturation of $\mathcal D_d(\rho)$. Within the OCP--RPA model, $\eta=2a_{\rm rel}^2Cd$ is independent of $\kappa$, so the concentration mainly shifts the crossover scales rather than the asymptotic slope.

\begin{figure*}[t]
\centering
\includegraphics[width=0.98\textwidth]{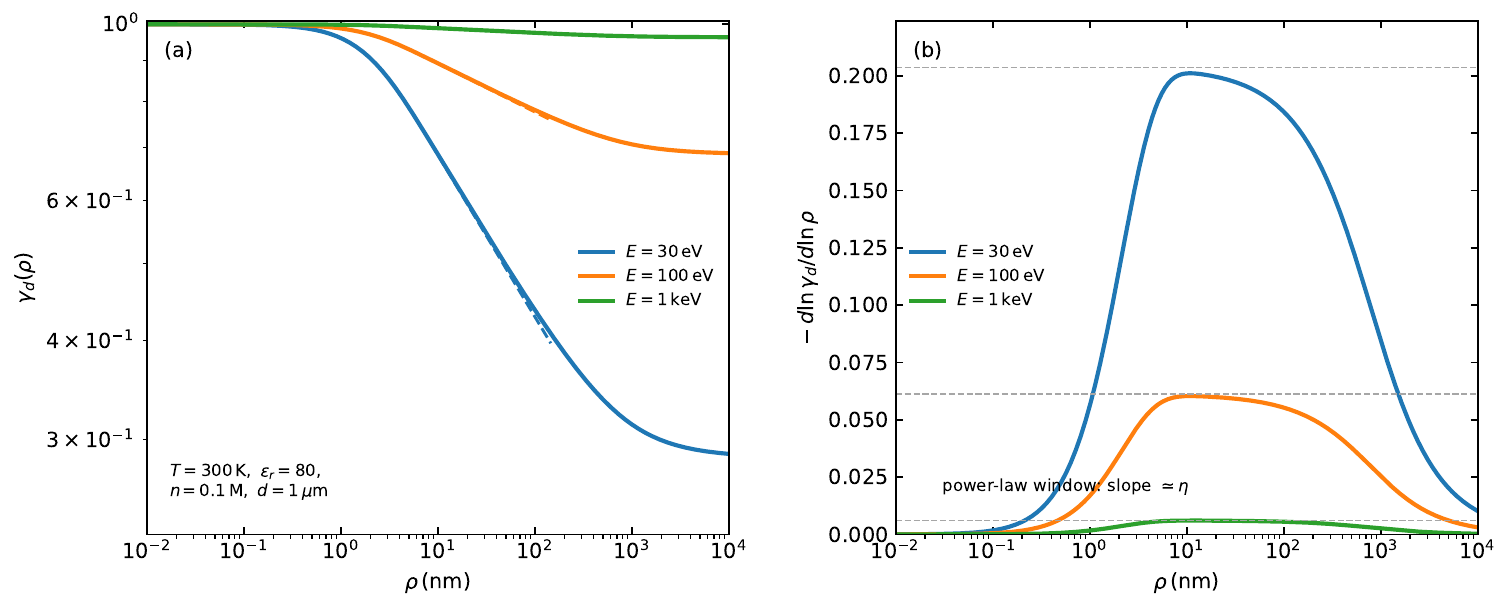}
\caption{Energy dependence of the finite-thickness mutual coherence at $T=300\,\mathrm K$, $\varepsilon_r=80$, $n=0.1\,\mathrm{mol\,L^{-1}}$, and $d=1\,\mu\mathrm m$. The electron energies are $30\,\mathrm{eV}$, $100\,\mathrm{eV}$, and $1\,\mathrm{keV}$. (a) Exact coherence envelopes (solid) and their intermediate power-law asymptotes (dashed). Reducing the energy increases $a_{\rm rel}=1/(\hbar v)$ and therefore both the algebraic exponent and the total coherence loss. (b) The local exponent approaches the corresponding value of $\eta=2a_{\rm rel}^{2}Cd$ in the intermediate window and returns to zero at large separation because of finite-thickness saturation.}
\label{fig:coherence_energy}
\end{figure*}

The dependence on the electron energy is shown in Fig.~\ref{fig:coherence_energy}. At fixed plasma parameters and slab thickness, the Debye crossover scale is unchanged, whereas the amplitude parameter $\eta=2a_{\rm rel}^{2}Cd=2Cd/(\hbar^{2}v^{2})$ increases as the electron velocity is reduced. The three curves therefore enter the algebraic regime at approximately the same transverse separation but acquire different slopes and different saturation plateaus. For the illustrative $d=1\,\mu\mathrm m$ slab, the exact plateau values are $\gamma_d(\infty)=0.2850$, $0.6862$, and $0.9629$ at $30\,\mathrm{eV}$, $100\,\mathrm{eV}$, and $1\,\mathrm{keV}$, respectively. The $30\,\mathrm{eV}$ curve satisfies $\mathcal D_d(\infty)>2$ and therefore possesses a finite $e^{-1}$ coherence radius, $\rho_c\simeq270\,\mathrm{nm}$, within the idealized static model, whereas the two higher-energy curves do not.

\begin{figure*}[t]
\centering
\includegraphics[width=0.98\textwidth]{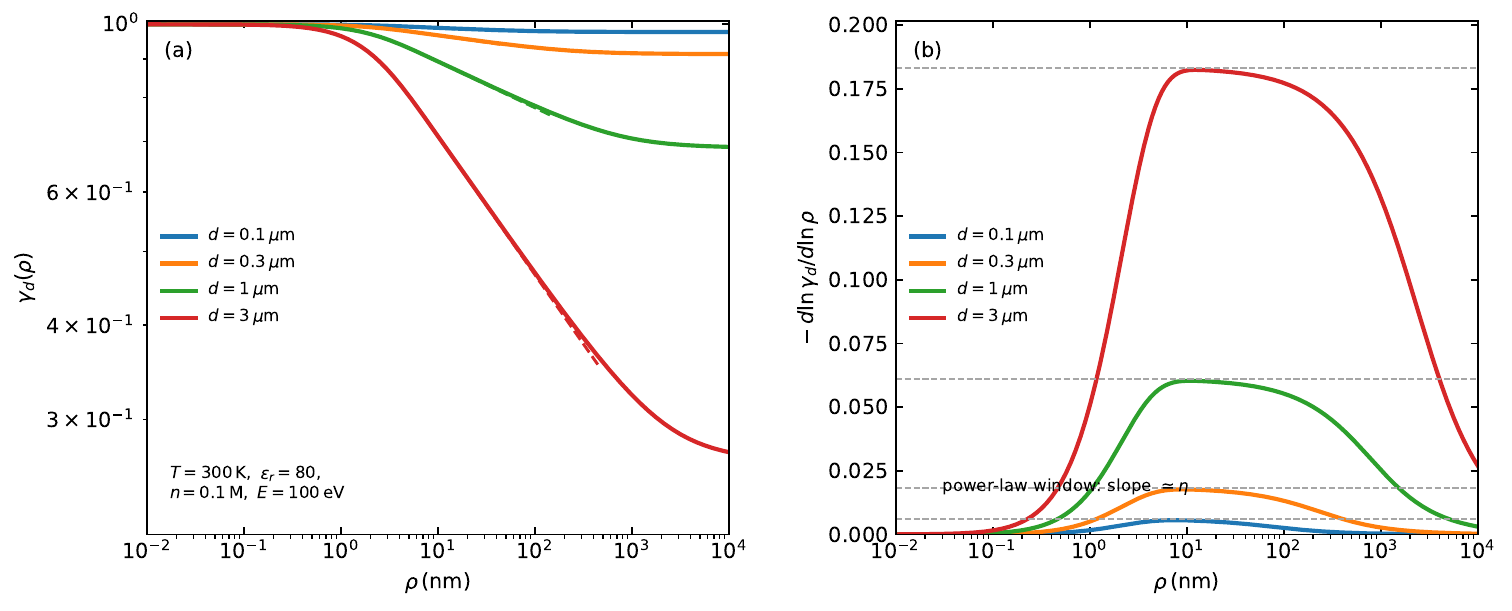}
\caption{Thickness dependence of the finite-slab mutual coherence for $T=300\,\mathrm K$, $\varepsilon_r=80$, $n=0.1\,\mathrm{mol\,L^{-1}}$, and $E=100\,\mathrm{eV}$. The thicknesses are $d=0.1$, $0.3$, $1$, and $3\,\mu\mathrm m$. (a) Exact coherence envelopes (solid) and intermediate power laws (dashed). Increasing $d$ both strengthens the decorrelation and delays the final saturation to larger transverse separation. (b) The local-exponent plateaus occur at $\eta=2a_{\rm rel}^{2}Cd$, demonstrating the predicted linear dependence of the algebraic exponent on thickness.}
\label{fig:coherence_thickness}
\end{figure*}

Figure~\ref{fig:coherence_thickness} isolates the effect of the slab thickness. Increasing $d$ has two simultaneous consequences: it enlarges the interval $\lambda_D\ll\rho\ll d$ in which the algebraic envelope can be observed, and it increases the exponent linearly, $\eta\propto d$. The exact asymptotic coherence levels for $d=0.1$, $0.3$, $1$, and $3\,\mu\mathrm m$ are $0.97660$, $0.91303$, $0.68616$, and $0.26419$, respectively. The last case has $\mathcal D_d(\infty)=2.662>2$ and hence admits a finite coherence radius, $\rho_c\simeq414\,\mathrm{nm}$; the thinner slabs remain above the $e^{-1}$ level for all $\rho$.

For an electron with a kinetic energy of $T_{\rm kin}=100\,\mathrm{keV}$, the relativistic wave number is $k=\sqrt{T_{\rm kin}(T_{\rm kin}+2mc^2)}/(\hbar c)\simeq1.70\times10^{12}\,\mathrm{m^{-1}}$. This value is used in the phase coupling term $1/(\hbar v)$, which automatically incorporates the relativistic correction. The latter will be discussed in more detail in Sec.~\ref{sec:relativistic}.

\subsection{Extension to a nonlocal dielectric response}

For a structured polar liquid, let $\epsilon_L(Q)$ denote the dimensionless static longitudinal relative permittivity, with $Q$ the fluctuation wave number. In SI units, the fluctuation--dissipation relation reads
\begin{equation}\label{eq:Kq_dielectric}
\widetilde K(Q)
=\frac{q_0^2k_{\mathrm B}T}{\varepsilon_0Q^2}
\left(1-\frac{1}{\epsilon_L(Q)}\right).
\end{equation}
If one isolates ionic fluctuations on top of a fixed background relative permittivity $\varepsilon_{b,r}$, the corresponding ionic dielectric function is defined relative to that background, and the prefactor becomes $q_0^2k_{\mathrm B}T/(\varepsilon_0\varepsilon_{b,r}Q^2)$. For the eikonal attenuation the relevant projection is
\begin{equation}\label{eq:line_integral_spectrum}
\int_0^\infty du\,K(u\mathbf n)
=\frac{1}{2}\int\frac{d^2Q_\perp}{(2\pi)^2}
\widetilde K(\mathbf Q_\perp,Q_\parallel=0),
\end{equation}
so that the transverse spectrum at zero longitudinal wave number is the relevant input. Oscillatory nonlocal response can produce corresponding structure in $K(r)$ and in the finite-slab phase structure function. Whether this leads to nonmonotonic mutual coherence depends on the full positive-definite spectral density and should be assessed for each microscopic dielectric model~\cite{bopp1996static,chen2025nonlocal}.

\section{Relativistic paraxial reduction and coherence theory}
\label{sec:relativistic}

Typical TEM beam energies are
$100$--$300\;\mathrm{keV}$, corresponding to velocities
$v\simeq0.55c$--$0.78c$. A relativistic treatment is therefore required
for a quantitatively consistent description of disorder-induced
mutual-coherence degradation. Relativistic formulations of electron
diffraction based on the Dirac equation and its paraxial reduction have
been considered previously~\cite{fujiwara1961relativistic,lentzen2023spin}.
Here we use the leading scalar paraxial reduction of the Dirac equation.
In this section $W({\bf r})$ denotes the electrostatic
\emph{potential energy} of the electron in the fluctuating Coulomb field.

The Dirac equation in Hamiltonian form reads
\begin{equation*}
i\hbar \partial_t \Psi
= \left( c\boldsymbol{\alpha}\cdot\hat{\mathbf p}
+ \beta mc^{2} + W(\mathbf r) \right) \Psi.
\end{equation*}
with the standard Dirac matrices
$\boldsymbol{\alpha}=\gamma^{0}\boldsymbol{\gamma}$ and
$\beta=\gamma^{0}$. For a stationary state
$\Psi({\bf r},t)=e^{-i{\cal E}t/\hbar}\Psi({\bf r})$, write the spinor as
$\Psi=(\varphi,\chi)^T$. The two coupled equations are
\begin{equation*}
\begin{aligned}
({\cal E}-W-mc^2)\varphi
&=c\,\boldsymbol\sigma\cdot\hat{\bf p}\,\chi,\\
({\cal E}-W+mc^2)\chi
&=c\,\boldsymbol\sigma\cdot\hat{\bf p}\,\varphi.
\end{aligned}
\end{equation*}
Eliminating the lower component gives $\chi=({\cal E}-W+mc^2)^{-1}c\,\boldsymbol\sigma\cdot\hat{\bf p}\,\varphi$. Because this operator does not commute with momentum, the exact eliminated equation contains gradients of $W$.
In the leading scalar paraxial approximation we keep the potential energy
$W$ to first order in the local dispersion relation but neglect its
gradients, which generate Darwin and spin--orbit terms. Then
\begin{equation*}
\left(({\cal E}-W)^2-m^2c^4\right)\varphi
\simeq
c^2\hat{\bf p}^{\,2}\varphi.
\end{equation*}
Equivalently,
\begin{equation*}
\nabla^2\varphi+
\frac{({\cal E}-W)^2-m^2c^4}{\hbar^2c^2}\,\varphi=0.
\end{equation*}
Expanding to first order in $W$ gives
\begin{equation*}
\nabla^2\varphi+
\left(
 k^2-\frac{2{\cal E}}{\hbar^2c^2}W({\bf r})
\right)\varphi=0,
\end{equation*}
where ${\cal E}=\gamma mc^2$, $p=\gamma mv$, $k=p/\hbar$, and $v=pc^2/{\cal E}$.

Separating the fast longitudinal phase, $\varphi({\boldsymbol\rho},z)=e^{ikz}\phi({\boldsymbol\rho},z)$, and assuming the usual forward paraxial condition $|\partial_z^2\phi|\ll k|\partial_z\phi|$, substitution into the linearized Helmholtz equation yields
\begin{equation*}
2ik\partial_z\phi
=
-\nabla_\perp^2\phi+
\frac{2{\cal E}}{\hbar^2c^2}W({\boldsymbol\rho},z)\phi.
\end{equation*}
Since ${\cal E}/(\hbar^2c^2k)=1/(\hbar v)$, this is equivalently $i\partial_z\phi=[-\nabla_\perp^2/(2k)+W/(\hbar v)]\phi$. After multiplication by $\hbar v$,
\begin{equation}\label{eq:paraxial}
i\hbar v\,\partial_z \phi
= \left(
-\frac{\hbar^{2}}{2\gamma m}\nabla_\perp^{2}
+ W(\mathbf r)
\right)\phi.
\end{equation}
This Schr\"odinger-like paraxial propagation equation is the
starting point for electron-wave propagation and multislice theories in
electron diffraction and transmission electron microscopy~\cite{cowley1957scattering,ishizuka1977multislice,ophus2017fast}.
The relativistic correction enters the transverse diffraction through
the effective mass $\gamma m$, while the scalar electrostatic potential
energy keeps unit coefficient in the $i\hbar v\partial_z$ form. Hence the
random eikonal phase is governed by the longitudinal velocity $v$.

In the straight-line eikonal approximation the accumulated random phase over a distance $d$ is $\Delta\phi=-(\hbar v)^{-1}\int_0^d dz\,W({\boldsymbol\rho},z)$, so that the effective coupling constant controlling the disorder-induced phase fluctuations is
\begin{equation}\label{eq:arel}
a_{\rm rel}=\frac{1}{\hbar v}
=\frac{1}{\hbar c\,\beta_v},
\qquad
\beta_v=\frac{v}{c}.
\end{equation}
This expression reduces to the standard nonrelativistic eikonal factor $1/(\hbar v)$ at low velocities and approaches the finite high-energy value $a_{\rm rel}\to a_\infty=1/(\hbar c)$ as $v\to c$.
The same result follows directly from the relativistic Hamilton--Jacobi
relation $({\cal E}-W)^2=m^2c^4+p^2c^2$: at fixed total energy,
$\delta p=-(\mathcal E/c^2p)W=-W/v$, and therefore
$\delta\phi=\hbar^{-1}\int dz\,\delta p$.

For statistically homogeneous Gaussian disorder, the finite-thickness formula in Eq.~\eqref{eq:Dphi_finite} retains its form after the replacement $a\to a_{\rm rel}=1/(\hbar v)$. In the ultrarelativistic limit the algebraic coherence regime acquires
the energy-independent exponent
\begin{equation}\label{eq:eta_infty}
\eta_\infty=\frac{2Cd}{\hbar^2c^2}.
\end{equation}
Thus, within the intermediate window $\lambda_D\ll\rho\ll d$,
\begin{equation}\label{eq:gamma_ur_power}
\gamma_d(\rho)\simeq
\exp(-\eta_\infty\gamma_{\rm E})
\left(\frac{\kappa\rho}{2}\right)^{-\eta_\infty}.
\end{equation}
As $v\to c$, the algebraic exponent approaches the finite value $\eta_\infty$ instead of vanishing. The finite-thickness plateau also has the finite limit $\gamma_d(\infty)\to\exp[-2Cd\,\mathcal H(\kappa d)/(\hbar^2c^2)]$. The power law remains an intermediate asymptote rather than a global one,
because finite slab thickness ultimately produces saturation for
$\rho\gtrsim d$.

In the weak-saddle regime the corresponding attenuation scale is
\begin{equation}\label{eq:ell_rel}
\begin{aligned}
\ell_{\rm rel,w}
&=\frac{1}{a_{\rm rel}^2C
\left(\ln(\kappa L_{\rm IR})+\gamma_{\rm E}
+E_1(\kappa L_{\rm IR})\right)}\\
&\simeq\frac{\hbar^2v^2}{C\ln(\kappa L_{\rm IR})}.
\end{aligned}
\end{equation}
For a fixed physical cutoff $L_{\rm IR}$, the ultrarelativistic limit is $\ell_{\rm rel,w}\to\ell_{\rm max}=\hbar^2c^2/[C\ln(\kappa L_{\rm IR})]$ within logarithmic accuracy. The finite-thickness plateau is obtained by
replacing $a$ with $a_{\rm rel}$ in Eq.~\eqref{eq:Dinf_OCP}; no universal
relativistic coherence radius follows unless the existence and
small-separation conditions of Eqs.~\eqref{eq:existence_condition} and
\eqref{eq:rhoc_conditional} are satisfied.

At $100\,\mathrm{keV}$ and $300\,\mathrm{keV}$ one has
$a_{\rm rel}/a_\infty=c/v\simeq1.82$ and $1.29$, respectively, where
$a_\infty=1/(\hbar c)$. The quasi-static Coulomb phase noise decreases
with accelerating voltage and approaches a finite high-energy limit.

\section{Conclusion}

The $1/r$ tail of the RPA potential correlator separates two aspects of propagation that are often described by a single disorder length. The disorder-averaged one-particle Green function is infrared sensitive: in an infinite static OCP its attenuation rate retains a logarithmic dependence on propagation distance, and a physical cutoff $L_{\rm IR}$ is needed to define $\ell_{\rm w}$. The formal strong saddle gives the momentum-independent asymptote $\ell_{\rm s}=(4\sqrt[3]{2}/3)G^{-1/3}$ within the same straight-line construction.

Mutual coherence behaves differently because the common long-wavelength phase cancels between two trajectories. For a finite slab, $\gamma_d(\rho)$ crosses from a small-separation regime to an algebraic envelope for $\lambda_D\ll\rho\ll d$ and finally to a nonzero plateau. A finite coherence radius appears only when that plateau lies below $e^{-1}$. When this threshold is crossed and $\kappa\rho_c\ll1$, the weak-saddle estimate reduces, up to logarithmic and order-unity factors, to $\rho_c\sim\lambda_D\sqrt{\ell_{\rm w}/d}$. This finite-size crossover is the main qualitative consequence of long-range Coulomb disorder for the two-path observable.

For an electron probe, the paraxial Dirac reduction gives $a_{\rm rel}=1/(\hbar v)$ and $\eta\to\eta_\infty=2Cd/(\hbar^2c^2)$ as $v\to c$. The resulting ionic dephasing is negligible for a $100\,\mathrm{nm}$ water-like layer at conventional TEM energies, but becomes stronger within the quasi-static model for thicker layers and lower-energy electrons. For liquid water, the sub-keV examples should be regarded as formal trends because inelastic and multiple scattering dominate practical transmission in that regime. The OCP provides the simplest realization of this mechanism; at Gaussian level the same long-wavelength correlator extends the result to dilute multicomponent electrolytes.

\appendix

\section{Gaussian transverse fluctuations about the rectilinear path}
\label{app:transverse_fluctuations}

Efimov's rectilinear-path approximation~\cite{efimov2015quantum} amounts to neglecting the
fluctuations around the straight trajectory in the functional integral.
Here we estimate explicitly the leading contribution of the two
transverse components. Starting from Eq.~\eqref{eq:G_avg_general}, write
$\boldsymbol\eta(u)=\eta_{\parallel}(u)\mathbf n+\boldsymbol\eta_\perp(u)$
with $\mathbf n\cdot\boldsymbol\eta_\perp(u)=0$. To isolate transverse
wandering we set $\eta_{\parallel}=0$. For $t=|u-u'|$ and
$\Delta\boldsymbol\eta_\perp=
\boldsymbol\eta_\perp(u)-\boldsymbol\eta_\perp(u')$, the separation
entering the correlator is
$R=[t^2+(s/k)|\Delta\boldsymbol\eta_\perp|^2]^{1/2}$ and
$K(R)=K(t)-(s/2k)J(t)|\Delta\boldsymbol\eta_\perp|^2
+O(\eta_\perp^4)$, where $J(t)=-K'(t)/t$.
The linear transverse variation vanishes by symmetry. The quadratic
part of the exponent relative to the straight path is therefore
\begin{equation*}
\begin{aligned}
\mathcal Q_\perp^{(2)}
={}&\frac{i}{2}\int_0^rdu\,
\dot{\boldsymbol\eta}_\perp^{\,2}(u)\\
&+\frac{m^2s^3}{4\hbar^4k^3}
\int_0^rdu\int_0^rdu'\,J(|u-u'|)\\
&\hspace{2.5em}\times
\left|\boldsymbol\eta_\perp(u)-\boldsymbol\eta_\perp(u')\right|^2.
\end{aligned}
\end{equation*}

For $r\to\infty$, endpoint effects do not contribute to the part of
$\ln G_k(r)$ that is extensive in $r$. Introducing longitudinal Fourier
modes, the finite combination that enters the quadratic form is
$\Delta J(q)=2\int_0^\infty dt\,J(t)[1-\cos(qt)]$. This difference is
important for the point-ion RPA correlator, for which $J(t)$ itself has
a short-distance singularity whereas $\Delta J(q)$ is finite. The
quadratic exponent becomes
\begin{equation*}
\mathcal Q_\perp^{(2)}
=\frac{1}{2}\int\frac{dq}{2\pi}
\left(
iq^2+
\frac{m^2s^3}{\hbar^4k^3}\Delta J(q)
\right)
\left|\boldsymbol\eta_\perp(q)\right|^2.
\end{equation*}

To make explicit how the imaginary unit in the free Fresnel kernel
enters the fluctuation correction, define
\[
X_q=
\frac{m^2s^3}{\hbar^4k^3}
\frac{\Delta J(q)}{q^2}.
\]
After normalization by the free transverse functional integral, each
mode contains the factor $1-iX_q$. For the two transverse components,
the part of the normalized determinant that is extensive in $r$ can be
written as
\[
\ln\frac{Z_\perp}{Z_{\perp,0}}
=
-r\int_{-\infty}^{\infty}\frac{dq}{2\pi}
\ln(1-iX_q)+o(r),
\]
where the retarded prescription fixes the branch continuously from
$X_q=0$. Since
\[
\ln(1-iX_q)
=
\frac{1}{2}\ln(1+X_q^2)
-i\arctan X_q,
\]
the imaginary unit does not disappear: the imaginary part of the
logarithm produces a fluctuation-induced phase shift, whereas its real
part produces additional attenuation. In particular, for $X_q\ll1$,
$\ln(1-iX_q)=-iX_q+X_q^2/2+O(X_q^3)$, so the correction linear in
$X_q$ is purely phase-like and the leading attenuation correction is
quadratic in the fluctuation coupling. Using the evenness of $X_q$,
the additional attenuation rate is therefore
\begin{equation}\label{eq:deltaGamma_transverse_general}
\delta\Gamma_\perp(s)
=
\int_0^\infty\frac{dq}{2\pi}
\ln\left(
1+
\left(
\frac{m^2s^3}{\hbar^4k^3}
\frac{\Delta J(q)}{q^2}
\right)^2
\right).
\end{equation}
The associated shift of the longitudinal saddle is of higher order in
the weak branch, so the leading transverse correction is obtained by
setting $s=1$ in Eq.~\eqref{eq:deltaGamma_transverse_general}.

For the OCP correlator in Eq.~\eqref{eq:K},
$J(t)=C[1-(1+\kappa t)e^{-\kappa t}]/t^3$. The corresponding
$\Delta J(q)$ can then be evaluated in closed form:
\begin{equation}\label{eq:DeltaJ_OCP_closed}
\begin{aligned}
\Delta J(q)
&=C\kappa^2 h\left(\frac{q}{\kappa}\right),\\
h(p)
&=\frac{1}{2}
\left[
(1+p^2)\ln(1+p^2)-p^2\ln p^2
\right].
\end{aligned}
\end{equation}
The limiting forms
$h(p)\simeq p^2[\ln(1/p)+1/2]$ for $p\ll1$ and
$h(p)\simeq\ln p+1/2$ for $p\gg1$ show that the determinant in
Eq.~\eqref{eq:deltaGamma_transverse_general} is convergent at both
limits. Introducing the dimensionless local coupling
$\chi=m^2C/(\hbar^4k^3)$, the weak-saddle expression reduces to
\begin{equation}\label{eq:deltaGamma_transverse_exact}
\delta\Gamma_\perp
=
\frac{\kappa}{2\pi}
\int_0^\infty dp\,
\ln\left[
1+\chi^2
\left(\frac{h(p)}{p^2}\right)^2
\right].
\end{equation}
For $\chi\to0$ its algebraic expansion is
\begin{equation}\label{eq:deltaGamma_transverse_weak}
\begin{aligned}
\delta\Gamma_\perp
&\simeq\frac{\kappa\chi^2}{2\pi}I_\perp
=0.590863\,\kappa\chi^2,\\
I_\perp&=\int_0^\infty dp\,
\left(\frac{h(p)}{p^2}\right)^2
=3.712501\ldots.
\end{aligned}
\end{equation}

To compare with the straight-line result, define
$\mathcal L=\ln(\kappa L_{\rm IR})+\gamma_{\rm E}
+E_1(\kappa L_{\rm IR})$, so that
$G=(m^2C/\hbar^4)\mathcal L$, $g=\chi\mathcal L$, and
$\Gamma_{\rm w}=G/k^2=k\chi\mathcal L$.
Equation~\eqref{eq:deltaGamma_transverse_weak} then gives
\begin{equation}\label{eq:transverse_ratio}
\frac{\delta\Gamma_\perp}{\Gamma_{\rm w}}
\simeq
0.590863\,
\frac{\kappa}{k}\,
\frac{\chi}{\mathcal L}
=
0.590863\,
\frac{\kappa}{k}\,
\frac{g}{\mathcal L^2}.
\end{equation}
Thus the leading transverse correction is parametrically small in the
controlled weak branch, where $g\ll1$ and $k\gg\kappa$. The two
conditions have distinct meanings: $g$ controls the longitudinal
saddle, while $\kappa/k$ controls transverse wandering on the Debye
scale. If the formal strong saddle is reached by lowering $k$ at fixed
medium parameters, both suppressions are progressively lost. This
supports the use of the rectilinear path for the weak branch but does
not provide a controlled continuation of Eq.~\eqref{eq:ell_strong}
into the regime $k\lesssim\kappa$.
\begin{acknowledgments}
The author thanks A.M. Satanin, L.N. Shchur, and P.E. Brandyshev for valuable discussions. The calculations were performed on the supercomputer facilities provided by NRU HSE.
\end{acknowledgments}

\bibliography{name}
\end{document}